# Quinone-based Switches for Candidate Building Blocks of Molecular Junctions with QTAIM and the Stress Tensor

Tianlv Xu[1], Lingling Wang[1], Yang Ping[1], Tanja van Mourik[2], Herbert Früchtl[2], Steven R. Kirk[*1] and Samantha Jenkins[*1]

[1]Key Laboratory of Chemical Biology and Traditional Chinese Medicine Research and Key Laboratory of Resource Fine-Processing and Advanced Materials of Hunan Province of MOE, College of Chemistry and Chemical Engineering, Hunan Normal University, Changsha, Hunan 410081, China

[2]EaStCHEM School of Chemistry, University of St Andrews, North Haugh, St Andrews, Fife KY16 9ST, Scotland, United Kingdom.

*email: samanthajsuman@gmail.com
*email: steven.kirk@cantab.net

The current work investigates candidate building blocks based on molecular junctions from hydrogen transfer tautomerization in the benzoquinone-like core of an azophenine molecule with QTAIM and the recently introduced stress tensor trajectory analysis. We find that in particular the stress tensor trajectories are well suited to describe the mechanism of the switching process. The effects of an Fe-dopant atom coordinated to the quinone ring, as well as F and Cl substitution of different ring-hydrogens, are investigated and the new QTAIM and stress tensor analysis is used to draw conclusions on the effectiveness of such molecules as molecular switches in nano-sized electronic circuits. We find that the coordinated Fe-dopant greatly improves the switching properties, both in terms of the tautomerisation barrier that has to be crossed in the switching process and the expected conductance behavior, while the effects of hydrogen substitution are more subtle. The absence of the Fe-dopant atom led to impaired functioning of the switch 'OFF' mechanism as well as coinciding with the formation of closed-shell H---H bond critical points that indicated a strained or electron deficient environment. Our analysis demonstrates promise for future use in design of molecular electronic devices.

# 1. Introduction

With silicon- and other semiconductor-based microelectronics approaching the limits of physically possible miniaturisation, the prospect of molecular electronics, with switches, transistors or memory elements the size of a single molecule, remains the holy grail of device building. Since the introduction of the term "molecular electronics" by Ratner in 2002[1], a considerable area of research has developed around molecule-sized switches. A recent review by Zhang *et al*[2] gives an overview of the breadth of the field. For the purposes of electronic components, molecular junctions are of particular interest, which have been reviewed by Komoto *et al*[3]. Major breakthroughs have recently been reported[4–7], but a workable strategy remains elusive.

We have found that azophenine, attached to a Cu(110) surface, can exist in two energetically identical states connected by hydrogen transfer from an amino to imino group and can be switched between those with the help of a Scanning Tunnelling Microscope (STM) tip [1]. Even more intriguingly, the energy difference of the most stable "*meta*" state and the metastable "*para*" structure has been calculated to be only 0.15 eV, with a conversion barrier of 0.42 eV.

For the purpose of molecular electronics 'free-standing' molecules are preferred to ones lying flat on a metal surface, since free-standing components could be attached to "molecular wires" made from, e.g., conducting polymers, linked together to form nano-sized integrated circuits. If electrical conductivity through the imino and amino groups differs, a switching event will lead to different electronic behaviour. Free-standing molecular components could also be used to build novel molecular sensors to control electronic properties such as resistivity based on the chemical environment.

In the gas phase however, we calculated the energy difference between the two states to be much larger, 0.55 eV (with the *para* geometry more stable) with a hydrogen transfer barrier of 0.79 eV, making the molecule much less desirable as a molecular electronics component. A family of molecular switches based on the same quinone core as azophenine, without the external phenyl groups, was recently proposed[8]. Initial calculations showed that a single iron or cobalt atom, coordinated to the central quinone ring, can significantly lower the energy difference and barrier height, similar to the effect of a metal surface. In this study, we will use 3-imino-6-methylenecyclohexa-1,4-diene-1,4-diamine, with and without F and Cl substituted at the ring hydrogens and with and without a coordinated Fe atom above the ring, see **Scheme 1**. In the manuscript it will be referred to as the quinone switch. This is a model system; for practical applications, the metal atom would have to be kept in place, *e.g.,* by being sandwiched between multiple rings or caged in by other molecules or functional groups.

A better understanding of the electronic structure would enable us to propose molecules with similar switching properties that might be more easily synthesised and networked than the model system proposed in Ref [8]. It would, for example, allow to predict the effect of different side groups attached to the quinone core, thus enabling fine-tuning of switching barrier and conductivity. This approach would include a formalism not in conventional Cartesian space but constructed from a 'phase-space' of the stress tensor trajectories that includes the most and least preferred directions of electronic charge density accumulation.

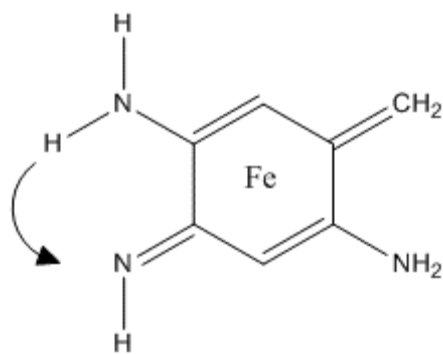

**Scheme 1:** Example switch molecule used in this study, without substituents. The arrow shows the hydrogen transfer involved in the switching process.

We propose to do this with the scalar[4–6] and vector-based[5,7] aspects of the quantum theory of atoms in molecules (QTAIM)[9] and the stress tensor formalism. In particular, we will use the QTAIM and stress tensor trajectory $\mathbb{T}_\sigma(s)$ formalisms in the stress tensor trajectory space $\mathbb{U}_\sigma$ to understand the functioning of the switch, previously used for benzene to present a new characterization of normal modes[10].

## 2. Theory and Methods
*2.1 The QTAIM and stress tensor BCP descriptors; ellipticity ε, the total local energy density $H(r_b)$ and stress tensor eigenvalue $\lambda_{3\sigma}$*

We use QTAIM[9] and the stress tensor analysis that utilizes higher derivatives of $\rho(\mathbf{r_b})$ in effect, acting as a 'magnifying lens' on the $\rho(\mathbf{r_b})$ derived properties of the wave-function. QTAIM allows us to identify critical points in the total electronic charge density distribution $\rho(\mathbf{r})$ by analyzing the gradient vector field $\nabla\rho(\mathbf{r})$. These critical points can be divided into four types of topologically stable critical points according to the set of ordered eigenvalues $\lambda_1 < \lambda_2 < \lambda_3$, with corresponding eigenvectors $\underline{e}_1$, $\underline{e}_2$, $\underline{e}_3$ of the Hessian matrix. In the limit that the forces on the nuclei become vanishingly small, an atomic interaction line (AIL)[11] becomes a bond-path, although not necessarily a chemical bond[12]. The complete set of critical points together with the bond-paths of a molecule or cluster is referred to as the molecular graph, with the constituent atoms being referred to as nuclear critical points (*NCPs*).

The ellipticity ε provides the relative accumulation of $\rho(\mathbf{r_b})$ in the two directions perpendicular to the bond-path at a *BCP*, defined as $\varepsilon = |\lambda_1|/|\lambda_2| - 1$ where $\lambda_1$ and $\lambda_2$ are negative eigenvalues of the corresponding and $\underline{e}_2$ respectively. It has been shown[13,14] that the degree of covalent character can be determined from the total local energy density $H(\mathbf{r}_b)$, defined as:

$$H(\mathbf{r}_b) = G(\mathbf{r}_b) + V(\mathbf{r}_b) \qquad (1)$$

In equation (1), $G(\mathbf{r}_b)$ and $V(\mathbf{r}_b)$ are the local kinetic and potential energy densities at a *BCP*, respectively. A value of $H(\mathbf{r}_b) < 0$ for the closed-shell interaction, $\nabla^2\rho(\mathbf{r}_b) > 0$, indicates a *BCP* with a degree of covalent character, and conversely $H(\mathbf{r}_b) > 0$ reveals a lack of covalent character for the closed-shell *BCP*. Throughout this work, we use the terminology '--' and '---' to refer to closed-shell *BCP*s which by definition always possess values of the Laplacian $\nabla^2\rho(\mathbf{r}_b) > 0$ but possess $H(\mathbf{r}_b) < 0$ or $H(\mathbf{r}_b) > 0$ respectively. Examples include the N10--H13/N10---H13 *BCP*, see **Figure 1** and **Figure 5(e)**. Conversely '-' always refers to shared-shell *BCP*s which by definition always possess values of the Laplacian $\nabla^2\rho(\mathbf{r}_b) < 0$ *and* $H(\mathbf{r}_b) < 0$, e.g. the double bond C2-C7 *BCP*, see **Figure 1** and **Figure 5(a)**.

The quantum stress tensor, $\sigma(\mathbf{r})$, is directly related to the Ehrenfest force by the virial theorem and so provides a physical explanation of the low frequency normal modes that accompany structural rearrangements[15–17]. In this investigation we will use Bader's definition[18,19] of the stress tensor. A diagonalization of the stress tensor, $\sigma(\mathbf{r})$, returns the principal electronic stresses. The stress tensor eigenvalue associated with the bond path, $\lambda_{3\sigma}$, has been associated with transition-type behavior in molecular motors[20].

*2.2 The stress tensor trajectory $\mathbb{T}_\sigma(s)$*

We will use the stress tensor eigenvectors $\{\underline{e}_{1\sigma}, \underline{e}_{2\sigma}, \underline{e}_{3\sigma}\}$, instead of the Cartesian coordinate frame, to track and characterize the changing orientation of the eigenvectors of the *BCP*s with respect to the functioning of the quinone switch. This is undertaken in terms of the opening, that is the 'ON' position, and closing of the switch, i.e. the 'OFF' position, both for the quinone switch with and without Fe-dopant, and with a fluorine atom in two different positions. The stress tensor eigenvectors $\underline{e}_{1\sigma}$ and $\underline{e}_{2\sigma}$ correspond to the most and least preferred directions of charge density accumulation $\rho(\mathbf{r}_b)$, and the $\underline{e}_{3\sigma}$ eigenvector is always directed along the bond path. To track the vector characteristics of the locations of the *BCP*s relative to the $\{\underline{e}_{1\sigma}, \underline{e}_{2\sigma}, \underline{e}_{3\sigma}\}$ framework of each *BCP* at the transition state (TS), throughout the ring-opening reaction, we visualize stress tensor $\mathbb{T}_\sigma(s)$ trajectories inside a stress tensor eigenvector projection space $\mathbb{U}_\sigma$ starting from the transition state. These stress tensor eigenvector trajectories $\mathbb{T}_\sigma(s)$ are constructed from the set of shifts $\mathbf{dr}(s)$, associated with steps *s*, where the parameter *s* is a sequence number of a given *BCP* in 3-D Cartesian space as an ordered set of vectors $\mathbf{dr'}(s)$ in the stress tensor eigenvector projection space $\mathbb{U}_\sigma$. We construct the trajectory $\mathbb{T}_\sigma(s)$ in the projection space $\mathbb{U}_\sigma$ by forming $\underline{\mathbf{e}}_{1\sigma}\cdot\mathbf{dr}(s)$, $\underline{\mathbf{e}}_{2\sigma}\cdot\mathbf{dr}(s)$ and $\underline{\mathbf{e}}_{3\sigma}\cdot\mathbf{dr}(s)$ where the transition state eigenvectors are used at each step down the Intrinsic Reaction Coordinate (IRC) starting from the transition state and ending at the corresponding minimum.

Additionally, for a given *BCP*, the stress tensor eigenvectors $\{\underline{\mathbf{e}}_{1\sigma}, \underline{\mathbf{e}}_{2\sigma}, \underline{\mathbf{e}}_{3\sigma}\}$ for that *BCP* at the transition state are used as the projection set for the entire trajectory $\mathbb{T}_\sigma(s)$. The corresponding trajectory length $\mathbb{L}_\sigma$ in the stress tensor eigenvector projection space $\mathbb{U}_\sigma$, is calculated as the sum:

$$\mathbb{L}_\sigma = \sum_{s=0} |\mathbf{dr'}(s+1) - \mathbf{dr'}(s)|. \qquad (2)$$

The trajectories $\mathbb{T}_\sigma(s)$ and the associated trajectory length $\mathbb{L}_\sigma$ can then be calculated for several of the reaction-pathways to highlight different behaviors. A longer trajectory length $\mathbb{L}_\sigma$ for a trajectory $\mathbb{T}_\sigma(s)$ implies a greater variation of a given *BCP*'s movement both in terms of direction in the stress tensor eigenvector projection space $\mathbb{U}_\sigma$ and the magnitude of the *BCP* shift, $\mathbf{dr'}(s)$. The trajectory length $\mathbb{L}_\sigma$ is calculated from the transition state ($s = 0$) to either the reverse or forward minimum in each case, see equation (**2**). If the magnitude, i.e. scalar length, of the *BCP* shift $\mathbf{dr'}(s)$ *and* the direction of $\mathbf{dr'}(s)$ are constant, then $\mathbb{L}_\sigma = 0$. Earlier, we used equation (**3**) to compare the values of the stress tensor trajectory length $\mathbb{L}_\sigma$ between the photochromism and fatigue path-ways *BCP*s of interest[21]. The corresponding real space lengths $l(s)$ of the $\mathbb{T}_\sigma(s)$ are calculated as the sum:

$$l = \sum_s |\mathbf{dr}(s)| \qquad (3)$$

## 3. Computational Details

Candidate structures for transition states were optimized with Gaussian 09[22] using DFT at the PBE0[23]/cc-pVTZ[24] level of theory, with Grimme's empirical 3-center dispersion correction with Becke-Johnson damping.[25,26] Gaussian's 'ultrafine' DFT integration grid was used for all calculations. These settings were retained in all subsequent calculations described below. The presence of exactly one imaginary vibrational frequency was confirmed for each transition state structure. All IRC calculations were performed with a step size of 0.03 Bohr, tight SCF convergence criteria and a termination criterion of energy gradient magnitude < $2 \times 10^{-4}$ Hartree/Bohr. The final calculated structures on each IRC path were then used as starting points for standard geometry optimizations, using the same method, basis set and DFT settings, to obtain precise local energy minimum structures to complete each full IRC path. For each generated structure representing a point on each computed IRC path, single-point SCF calculations were performed with the aforementioned DFT settings and basis set, with additional stricter convergence criteria; < $10^{-10}$ RMS change in the density matrix and < $10^{-8}$ maximum change in the density matrix. These calculations yielded the wave-functions needed for QTAIM analysis.

Calculations of the molecular graphs and critical point properties were performed using AIMAll[27]: all molecular graphs were additionally confirmed to be free of non-nuclear attractor critical points.

## 4. Results and Discussion

*4.1 Factors affecting the performance of the quinone switch from QTAIM and stress tensor*

The goal of this study is to understand the shape of the switching barrier and pathways of electric conductance with the help of QTAIM analysis and stress tensor analysis. Our model system is the molecule shown in **Scheme 1**. The intended conduction pathway goes through the imino or amino-nitrogen (atom N11) and the double-bonded carbon (C7). The effect of substitution of hydrogens attached to the ring carbons C3 and C6 by F and Cl atoms (effects of the latter shown in the **Supplementary Materials S3**) was investigated. Geometric descriptions using the notation "UP/DOWN" (attached to C3 or C6, respectively) and "*North*/*South* path" (C5-C4-C3-C2/C5-C6-C1-C2) in the further text are based on the orientation of **Figure 1**.

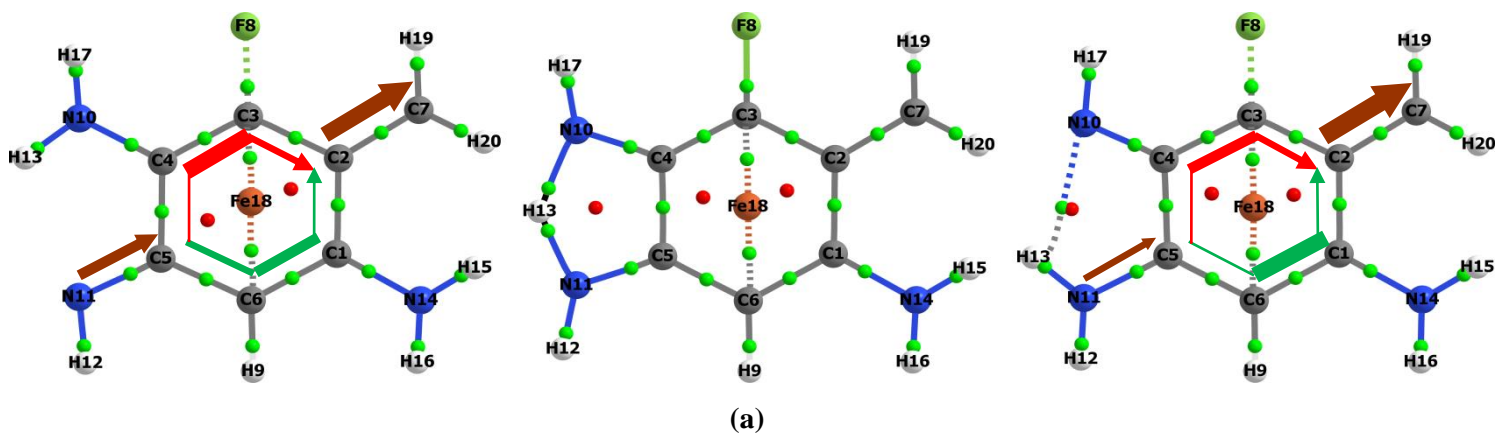

(a)

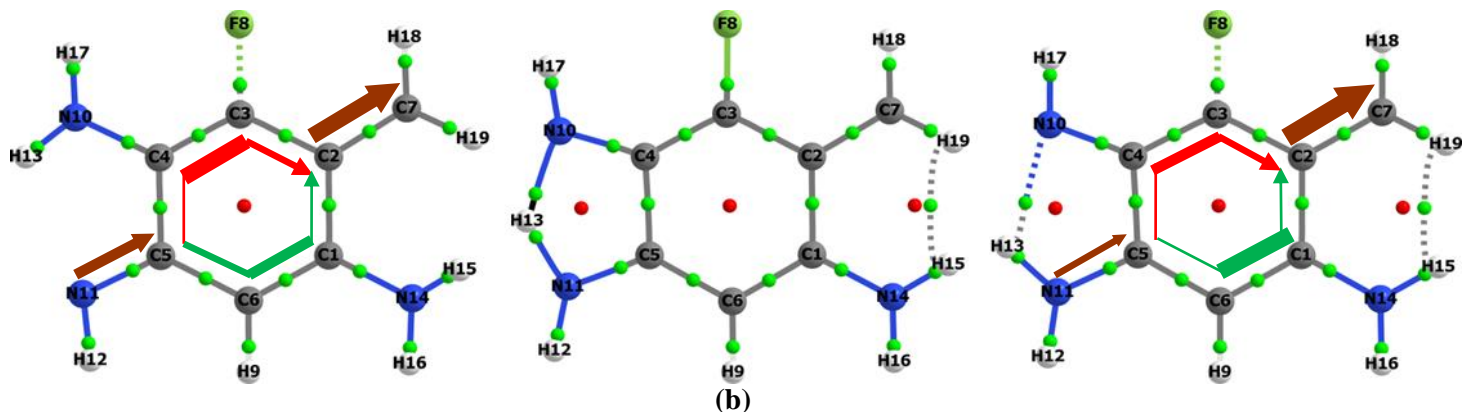

**Figure 1.** Snapshots of the molecular graphs of the F decorated quinone switch is shown doped with Fe in sub-figure **(a)** and undoped in sub-figure **(b)**. The left (the reverse minimum ($r$)) and right panel (forward minimum ($f$)) correspond to the 'ON' position and 'OFF' positions of the switch, respectively, with the middle panel corresponding to the transitions state (TS) of the IRC. The thickness of the red, green and brown arrows represent the magnitudes of the ellipticity ε of the *North* pathway (C5-C4-C3-C2), *South* pathway (C5-C6-C1-C2), the N11-C5 *BCP* ('entry' pathway) and C2-C7 *BCP* ('exit' pathway) respectively.

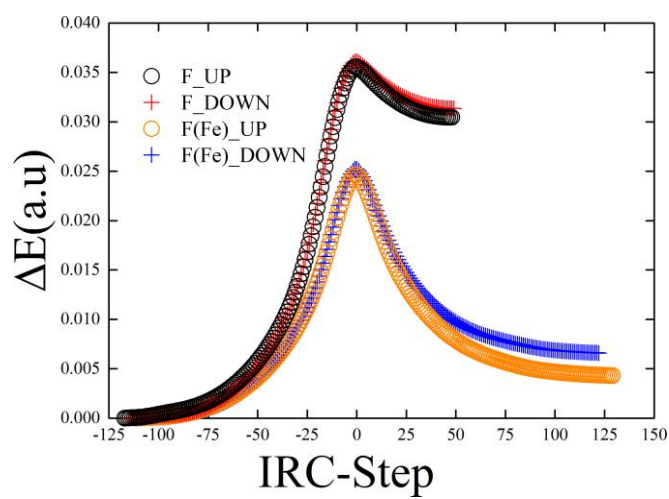

**Figure 2.** The relative energy ΔE of the F (Fe) UP and F (Fe) DOWN substituted reaction pathways with Fe-dopant *NCP* are indicated by the black circles and red pluses respectively. The corresponding ΔE plots without Fe (F UP and F DOWN) are indicated by the orange circles and blue pluses respectively. The corresponding results for the undecorated and Cl decorated quinone switch can be found in the **Supplementary Materials S1** and **Supplementary Materials S3** respectively. See **Figure 1** for the positions of the F (UP) and F (DOWN) locations respectively.

It is very evident that the presence of the Fe-dopant *NCP* is effective in yielding the desired symmetrical shape of the barrier in the quinone switch, with the positioning of the F either in the UP or DOWN position as secondary, see the relative energy ΔE plots in **Figure 2**. In this investigation we vary two variables: the presence of the Fe-dopant *NCP* and the position of an F *NCP*; either in the UP or DOWN position. In the molecular graphs presented in **Figure 1(a-b)** we show the F decorated quinone switch in the UP position, i.e. forming a bond-path with the C3 atom. The results for the DOWN position for the F NCP, which then corresponds to forming a bond-path with the C6 *NCP*, are given in the **Supplementary Materials S2**. The most likely conductance pathways of the quinone switch, on the basis of higher values of the ellipticity ε, are indicated by arrows that are scaled with the magnitude of the ellipticity ε of the corresponding *BCP*s in the molecular graphs presented in **Figure 1**. Therefore, we see that the reverse (negative IRC coordinate) and forward (positive IRC coordinate) directions correspond to the 'ON' and 'OFF' positions of the quinone switch respectively.

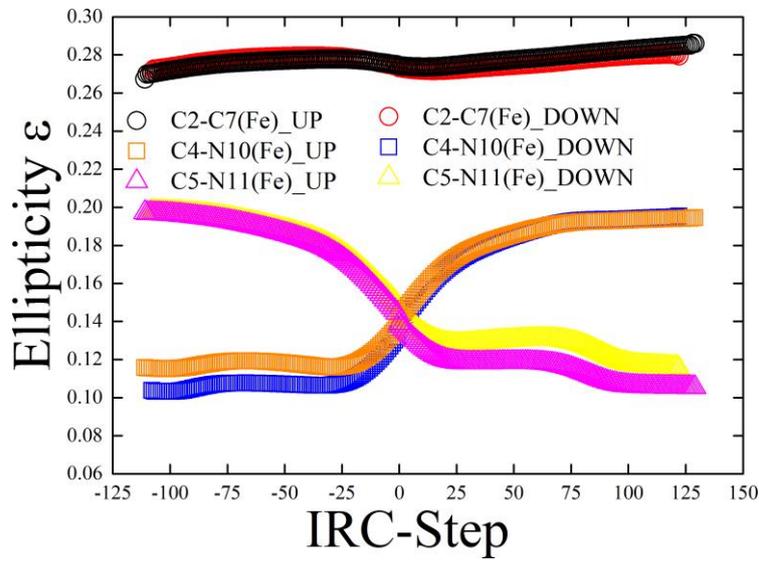

(a)

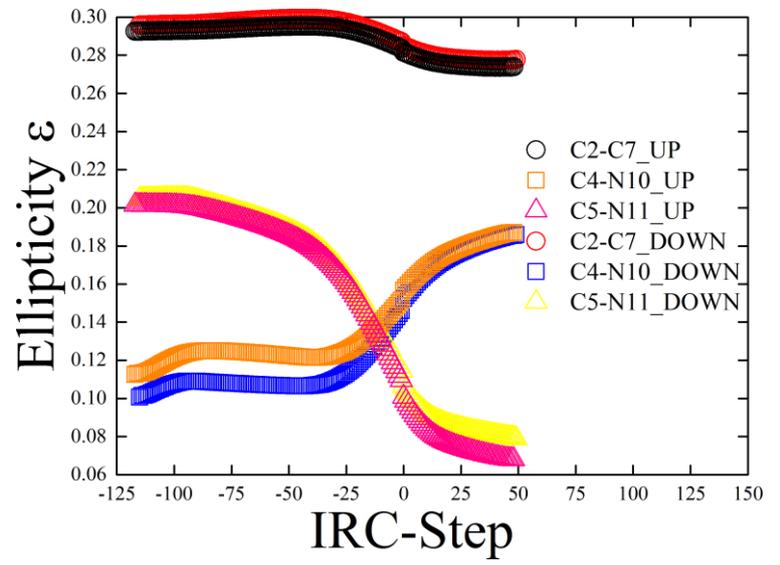

(b)

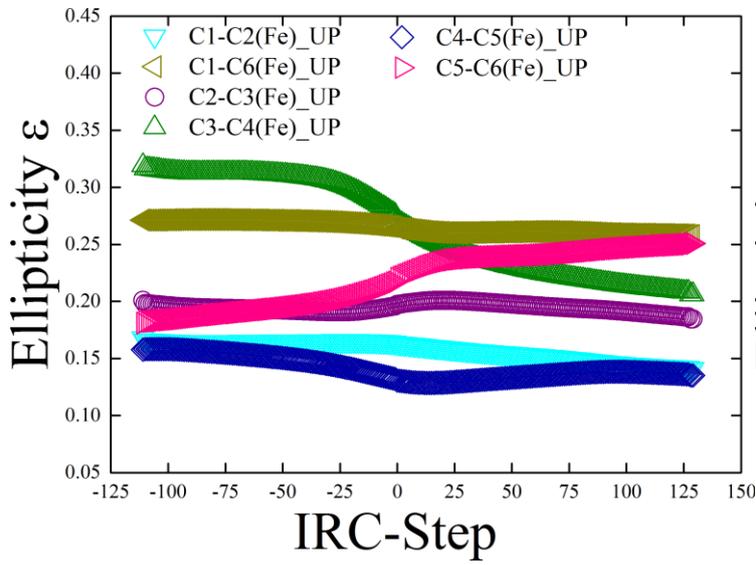

(c)

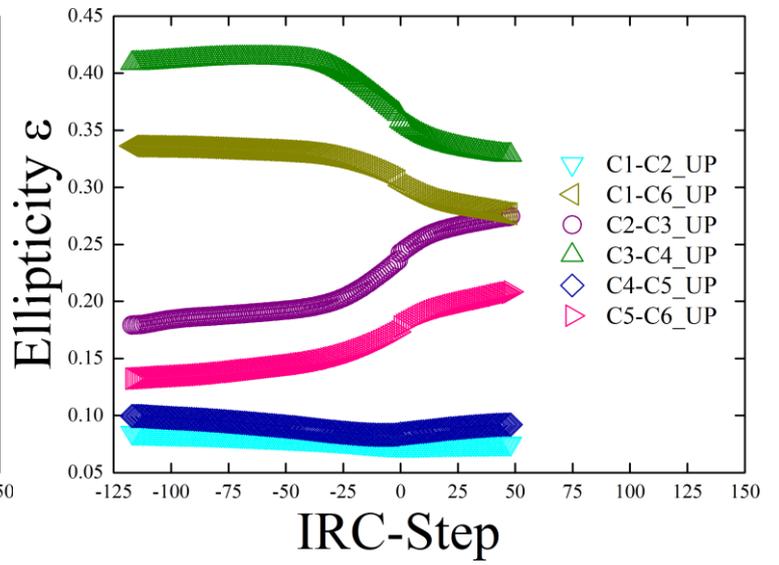

(d)

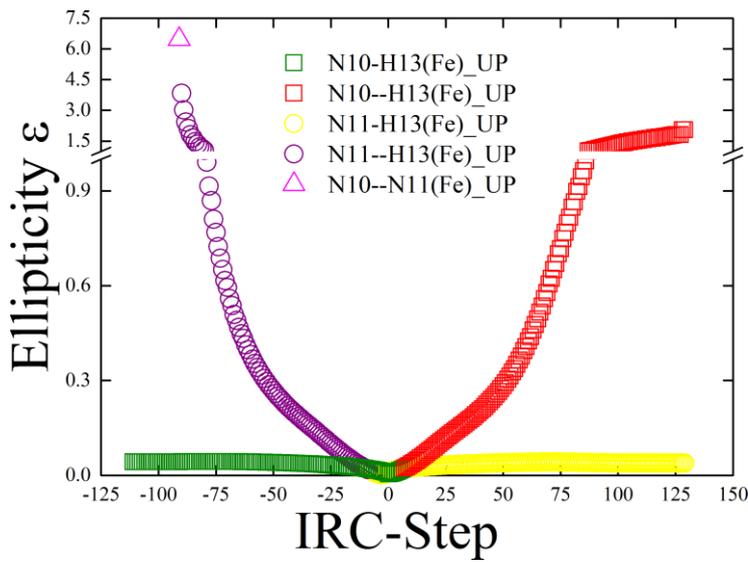

(e)

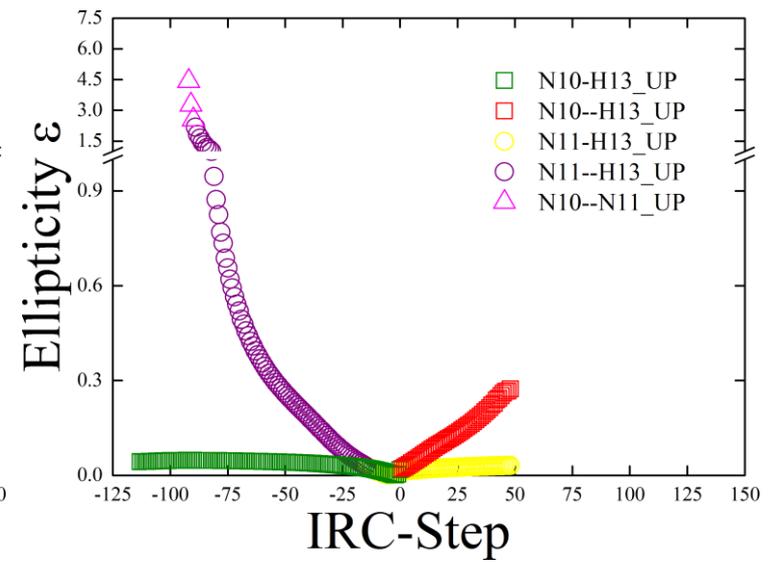

(f)

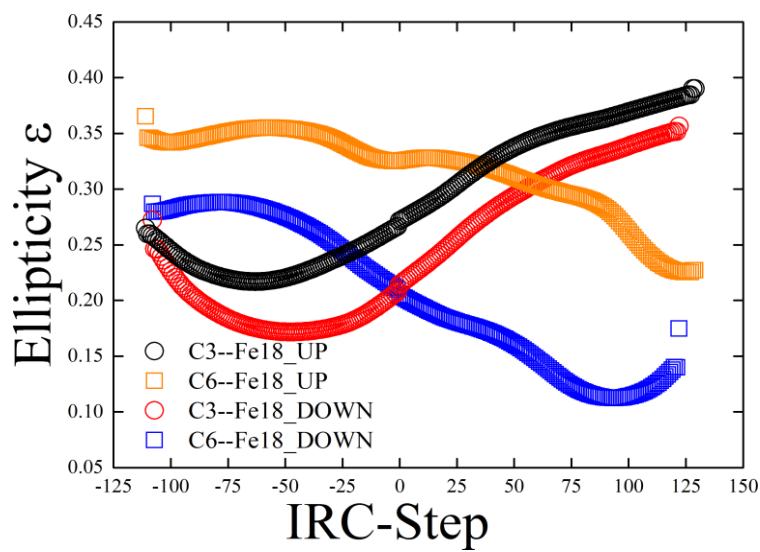

**(g)**

**Figure 3.** The variation of ellipticity ε with the IRC for the C2-C7 *BCP*, C4-N10 *BCP* and C5-N11 *BCP* of F with Fe and no Fe is presented in sub-figure **(a)** and **(b)** respectively. The corresponding ellipticity ε plots for the C-C *BCP*s, in **(c)** and **(d)** N-H *BCP*s in **(e)** and **(f)** C3--Fe18 *BCP* and C6--Fe18 *BCP* in **(g)** of F with Fe and no Fe are shown in sub-figure (**c-g**) respectively. See the caption of **Figure 2** for further details. For the locations of the *BCP*s and the atom labels refer to **Figure 1**.

The reverse direction corresponds to the 'ON' position for the Fe-doped switch with the C5-N11 *BCP* as the 'entry' pathway as indicated by the very thick brown arrow in the left panel of **Figure 1(a)**. The value of the ellipticity ε of the C5-N11 *BCP* is high enough to correspond to double bond character which implies higher conductivity for shared-shell *BCP*s. In the forward (*f*) direction, i.e. the 'OFF' position, the ellipticity ε of the Fe-doped switch of the C5-N11 *BCP* reduces to the extent of being a single bond, i.e. less favorable for conduction, indicated by the thin brown arrow in the right hand panel of **Figure 1(a)**. It can be seen that for the undoped quinone switch the C5-N11 *BCP* ellipticity ε falls to a lower value than for the Fe-doped switch, indicating that the switch is even less favorable for conduction in the 'OFF' position, see **Figure 3(b)**. The ellipticity ε of the C2-C7 *BCP* (the 'exit' pathway) for both the Fe-doped and undoped quinone switch remains high throughout both reactions consistent with the presence of a double bond, see **Figure 3**. Consistency is found from comparison with the total local energy density $H(\mathbf{r_b})$ where more negative values correspond to stronger bonds, see **Figure 4(a-b)** and correlates with higher ellipticity ε values, seen by comparison with **Figure 3(a-b)**.

Comparison of the stress tensor eigenvalue $\lambda_{3\sigma}$ of the Fe-doped and undoped C5-N11 *BCP* (the 'entry' pathway) shows more negative values of $\lambda_{3\sigma} < 0$ for the undoped switch showing *BCP* instability, seen by comparing **Figure 5(a)** with **Figure 5(b)**.

The effect of doping the quinone switch with Fe is to average out the values of the ellipticity ε of the ring C-C *BCP*s relative to the undoped switch; this will result in less favoring of the *North* pathway (C5-C4-C3-C2) compared with the *South* pathway (C5-C6-C1-C2), see **Figure 1** and compare **Figure 3(c)** with **Figure 3(d)**. Examination of the corresponding plots of $H(\mathbf{r_b})$, see **Figure 4(c-d)** as well as examination of the stress tensor eigenvalue $\lambda_{3\sigma}$, see **Figure 5(c-d)**, confirms this finding.

We see that for the proton transfer from N10 to N11 there is an exchange of ellipticity ε values (high to low in the bond path to the nitrogen gaining the transferred proton and vice versa for the other nitrogen) and also an exchange of chemical character, see **Figure 3(e-f)**. This exchange occurs between the shared-shell N10-H13 *BCP* to the closed-shell H13--N10 *BCP* from the reverse (*r*) through to the forward minimum (*f*) and for the closed-shell N11--H13 *BCP* to the shared-shell H13-N11 *BCP*, see **Figure 4(e-f).** This effect is also observed for the stress tensor eigenvalue $\lambda_{3\sigma}$, see **Figure 5(e-f)**. If the Fe-doped quinone switch is in the 'ON' position, there are no closed-shell *BCP*s present in the molecular graph; however this is not the case for the 'OFF' position of the switch, see the left panel of **Figure 1(a)**. Furthermore, for the undoped quinone switch in the 'OFF' position there are two closed-shell *BCP*s (H13--N10 *BCP* and H---H *BCP*)[28,29]. The presence of an H---H *BCP* suggests that the electronic environment for the undoped switch is more strained in the 'OFF' position than in the Fe-doped switch, compare the right panels of **Figure 1(a)** and **Figure 1(b)**. Assuming that the Fe-doped quinone switch shows better switching characteristics than the undoped switch, we now consider the factors affecting whether the UP or DOWN F decorated switch is preferred. We will examine this with respect to the C3-Fe18 *BCP* (UP) and the C6-Fe18 *BCP* (DOWN), see **Figure 3(g)**. Both of the C3-F18 *BCP* (UP) and the C6-Fe18 *BCP* (UP) possess greater ellipticity ε values than either of the C3-Fe18 *BCP* (DOWN) and C6-Fe18 *BCP* (DOWN). Therefore, this can explain why the switching characteristics of the Fe-doped molecule show a clear preference for the UP position of the F compared with the molecular graph with the F in the DOWN position, see **Figure 1** and **Figure 3(g)**. This is consistent with the observation that the C3-Fe18 *BCP* (UP) has more negative values of the total local energy density $H(\mathbf{r_b})$ than the corresponding C6-Fe18 *BCP* (DOWN), see **Figure 4(g).** Additional confirmation is found from presence of $\lambda_{3\sigma} > 0$ for the C3-Fe18 *BCP* (UP) and $\lambda_{3\sigma} < 0$ for the C6-Fe18 *BCP* (DOWN), see **Figure 5(g)**. Qualitatively similar results are seen for the H (i.e. unsubstituted) and Cl substituted molecules, with a DOWN variant preferred for Cl substituted quinone switches, see **Supplementary Materials S1** and **Supplementary Materials S3** respectively.

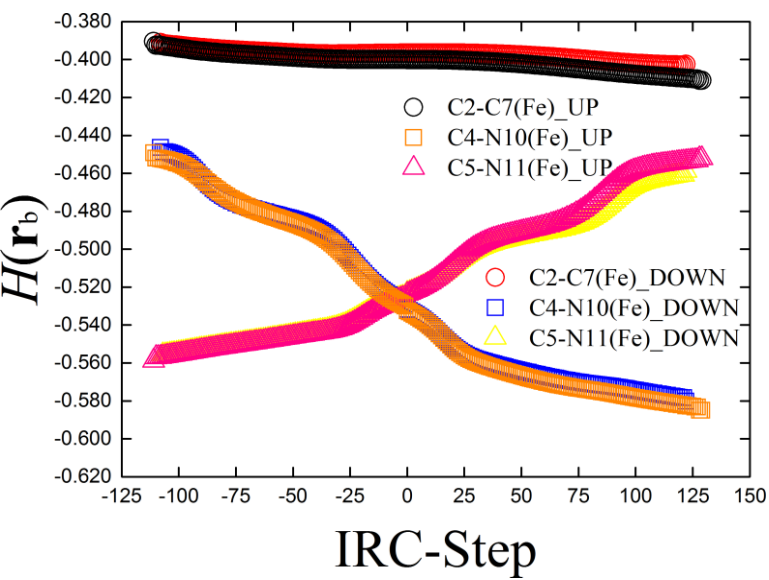
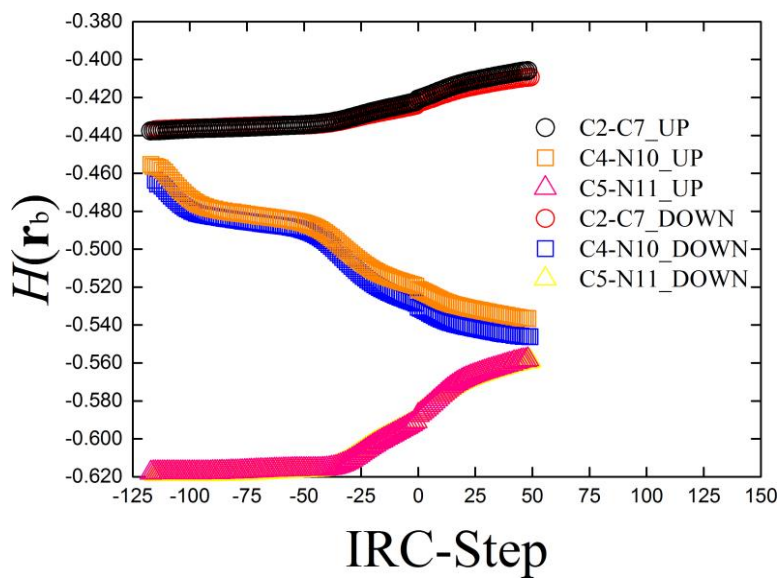

**(a)**             **(b)**

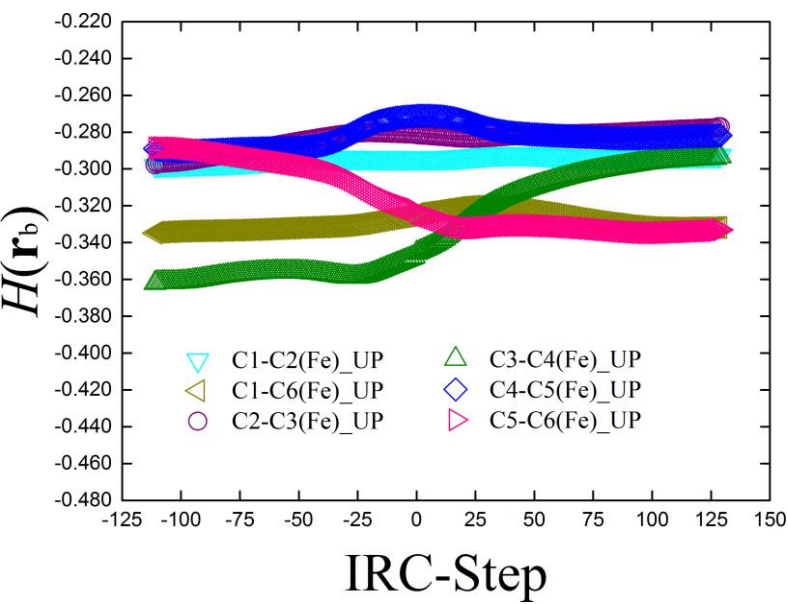
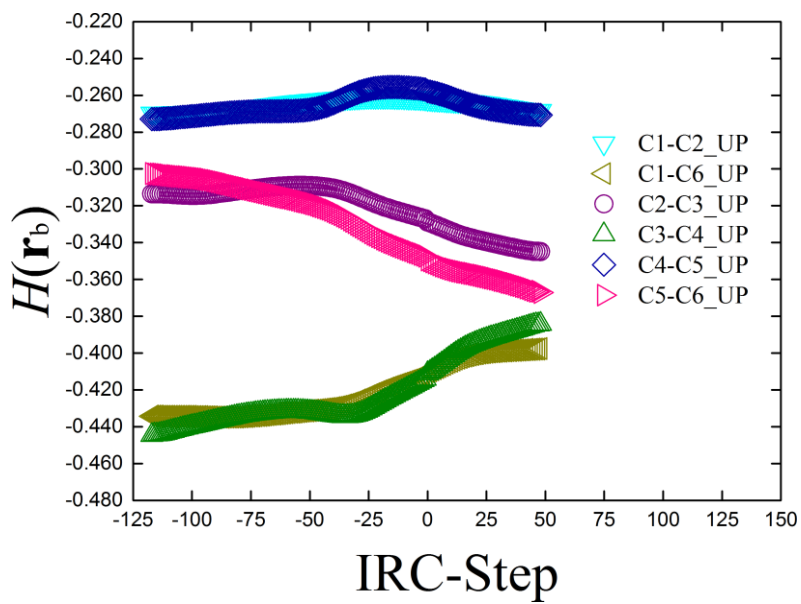

**(c)**             **(d)**

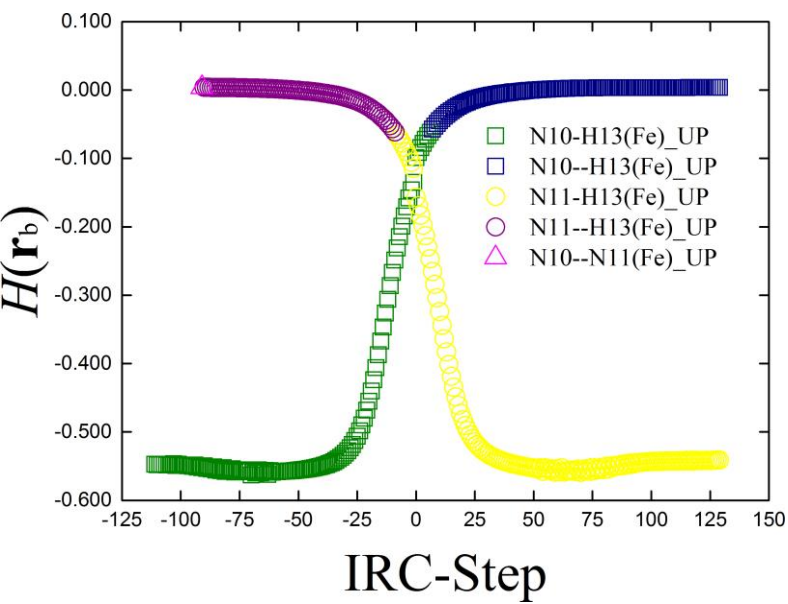
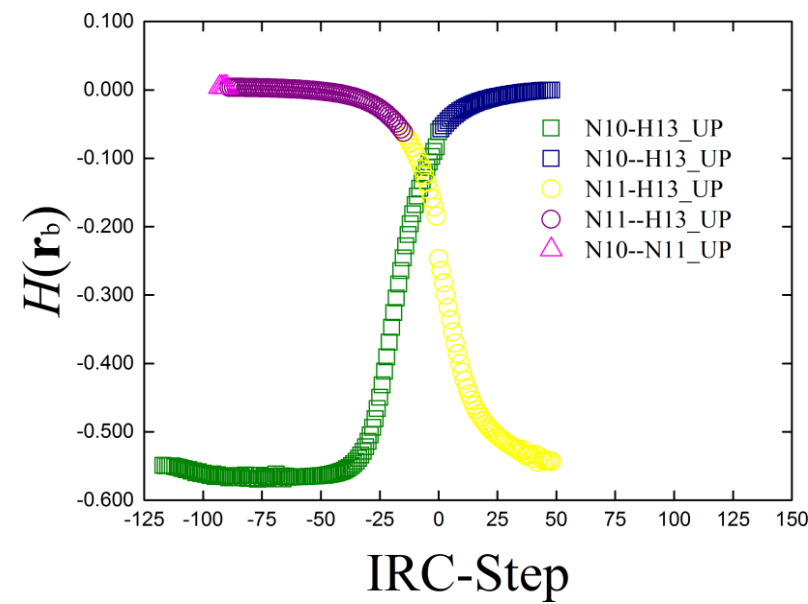

**(e)**             **(f)**

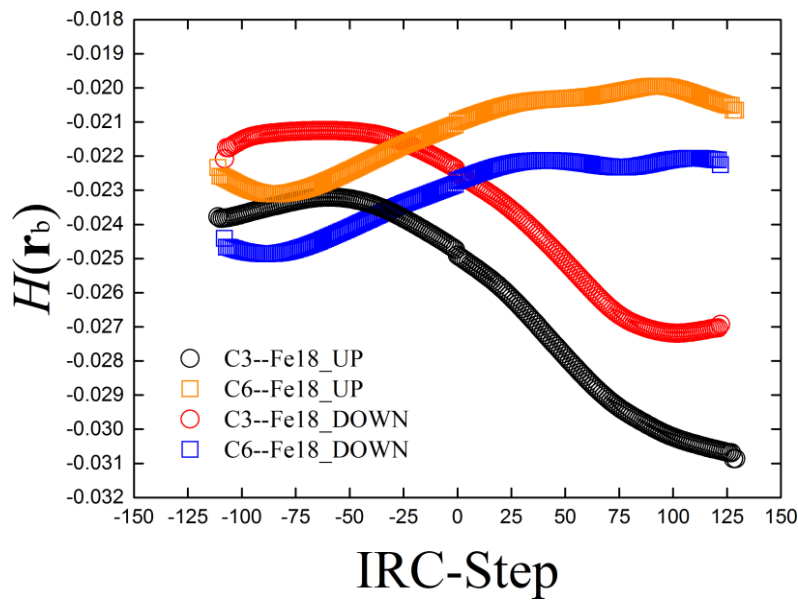

**(g)**

**Figure 4**. The variation of the total local energy density $H(\mathbf{r_b})$ with the IRC for C2-C7 *BCP*, C4-N10 *BCP* and C5-N11 *BCP* of F with Fe and no Fe are presented in sub-figure **(a)** and **(b)** respectively. The corresponding $H(\mathbf{r_b})$ plots for the C-C *BCP*s, N-H *BCP*s and C3--Fe18 *BCP* and C6--Fe18 *BCP* of F with Fe and no Fe are shown in sub-figure (**c-g**) respectively. See the caption of **Figure 3** for further details

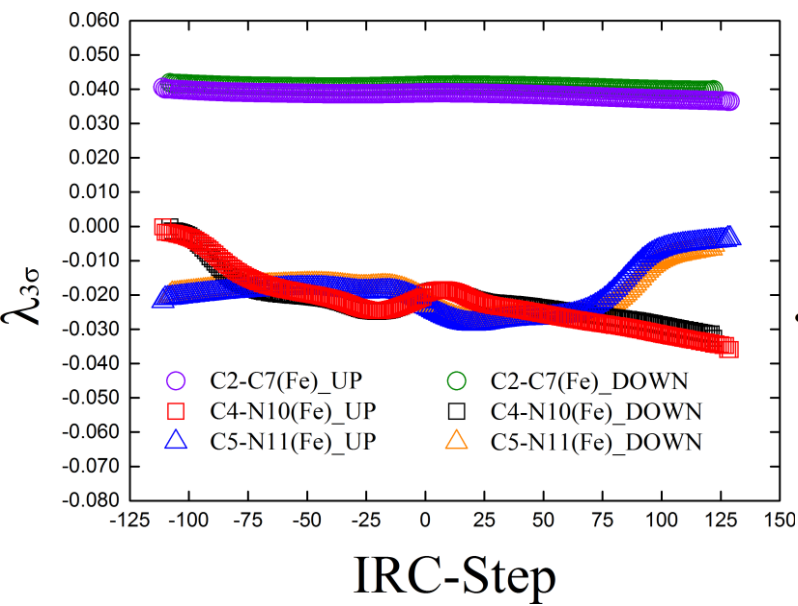

**(a)**

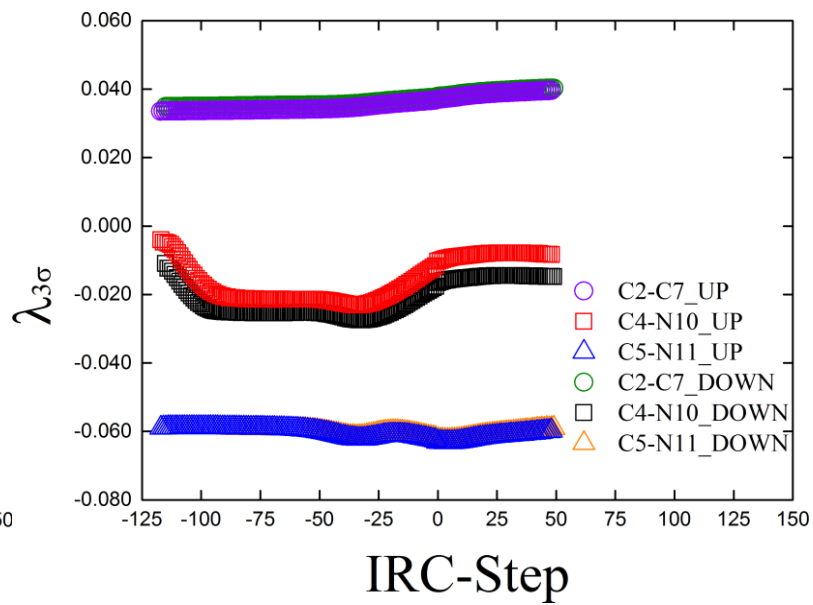

**(b)**

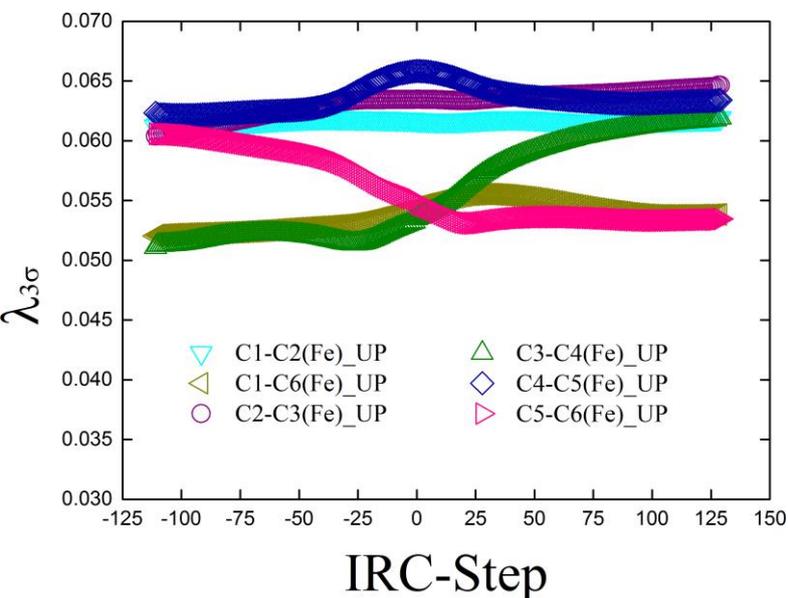

(c)

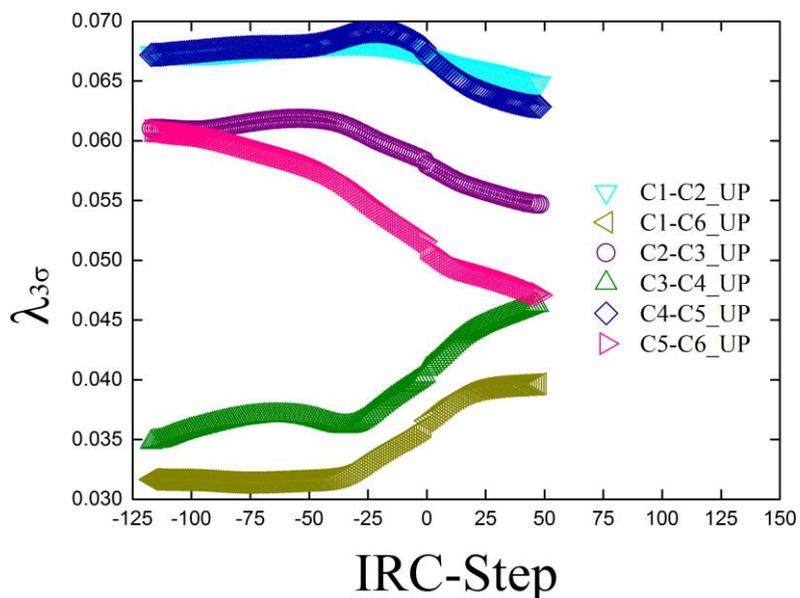

(d)

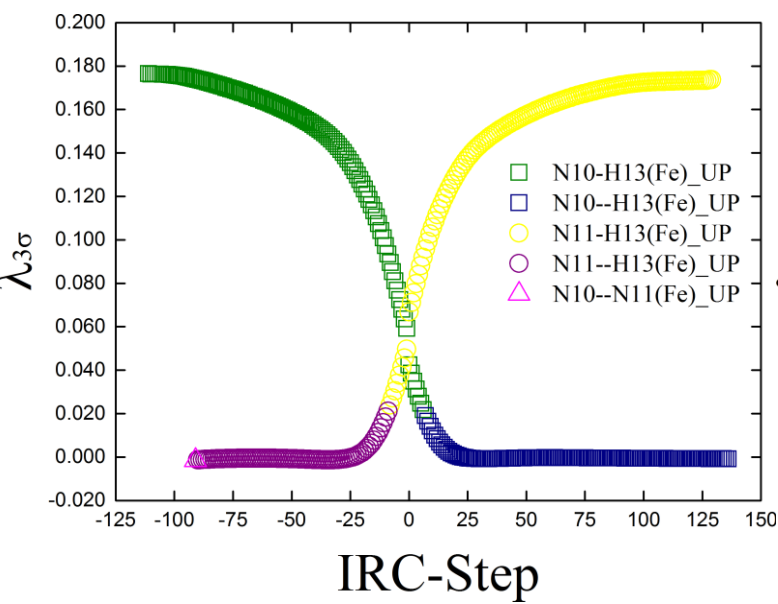

(e)

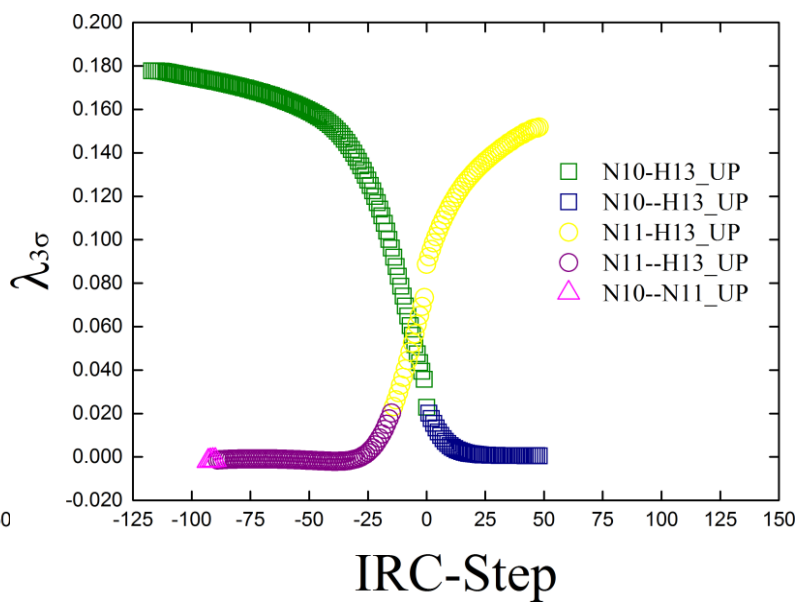

(f)

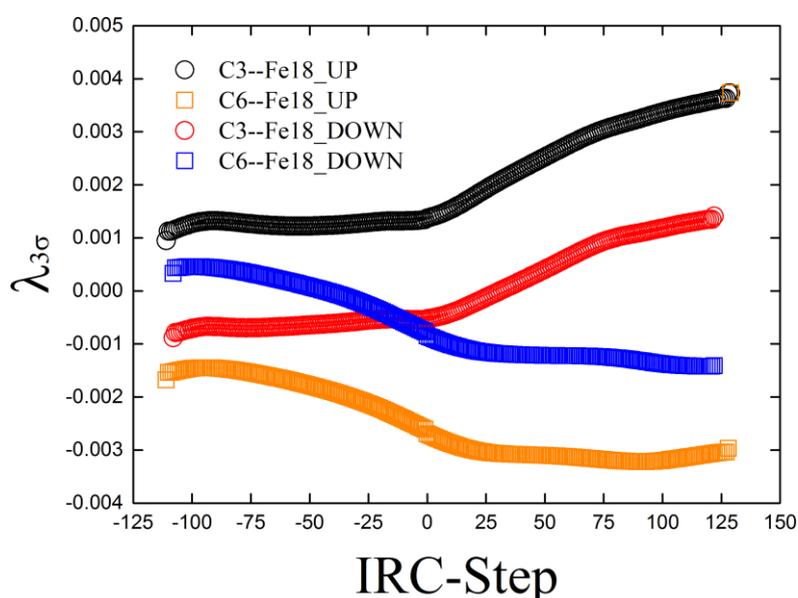

(g)

**Figure 5**. The variation of the stress tensor eigenvalue $\lambda_{3\sigma}$ with the IRC for C2-C7 *BCP*, C4-N10 *BCP* and C5-N11 *BCP* of the F substituted molecule with Fe and no Fe are presented in sub-figure **(a)** and **(b)** respectively. The corresponding $\lambda_{3\sigma}$ plots for the C-C *BCP*s, N-H *BCP*s and C3--Fe18 *BCP* and C6--Fe18 *BCP* of the F substituted molecule with Fe and no Fe are shown in sub-figure (**c-g**) respectively. See the caption of **Figure 3** for further details.

## 4.2 The functioning of the quinone switch explained by the stress tensor trajectory $\mathbb{T}_\sigma(s)$

The main goal of the stress tensor trajectory $\mathbb{T}_\sigma(s)$ analysis is to understand the functioning of the quinone switch in terms of the factors impairing the 'ON' and 'OFF' functioning of the switch in the stress tensor eigenvector projection space $\mathbb{U}_\sigma$. This is possible because the stress tensor trajectory $\mathbb{T}_\sigma(s)$ can be used to visualize the 'ON' or 'OFF' switching *process* in the stress tensor eigenvector projection space $\mathbb{U}_\sigma$. This is achieved by tracking the motion of a *BCP* starting from the transition state and terminating at either the reverse (*r*) or forward (*f*) minimum, corresponding to the 'ON' and 'OFF' positions respectively, see **Figure 1**. The corresponding trajectories $\mathbb{T}_\sigma(s)$ for the H and Cl decorated quinone switch show similar trends and can be found in the **Supplementary Materials S1** and **Supplementary Materials S3,** respectively.

To capture the locations of the *BCP*s relative to the $\{\underline{\mathbf{e}}_{1\sigma}, \underline{\mathbf{e}}_{2\sigma}, \underline{\mathbf{e}}_{3\sigma}\}$ framework along the reaction pathways from the transition state towards the reverse (*r*), 'ON' and forward (*f*), 'OFF' positions, we visualize the stress tensor trajectories $\mathbb{T}_\sigma(s)$ inside a stress tensor eigenvector projection space $\mathbb{U}_\sigma$ for the undoped and Fe-doped quinone switches (see theory section 2.2) and **Figures 6-8**. The corresponding lengths $\mathbb{L}_\sigma$ in $\mathbb{U}_\sigma$ space and lengths *l* of the trajectories $\mathbb{T}_\sigma(s)$ in real space are presented in **Table 1** and **Table 2**. The maximum projections $\{(\underline{\mathbf{e}}_{1\sigma}\cdot\mathbf{dr})_{max}, (\underline{\mathbf{e}}_{2\sigma}\cdot\mathbf{dr})_{max}, (\underline{\mathbf{e}}_{3\sigma}\cdot\mathbf{dr})_{max}\}$ in the stress tensor eigenvector projection space $\mathbb{U}_\sigma$ for the undoped and Fe-doped quinone switches are presented in **Table 3** and **Table 4,** respectively.

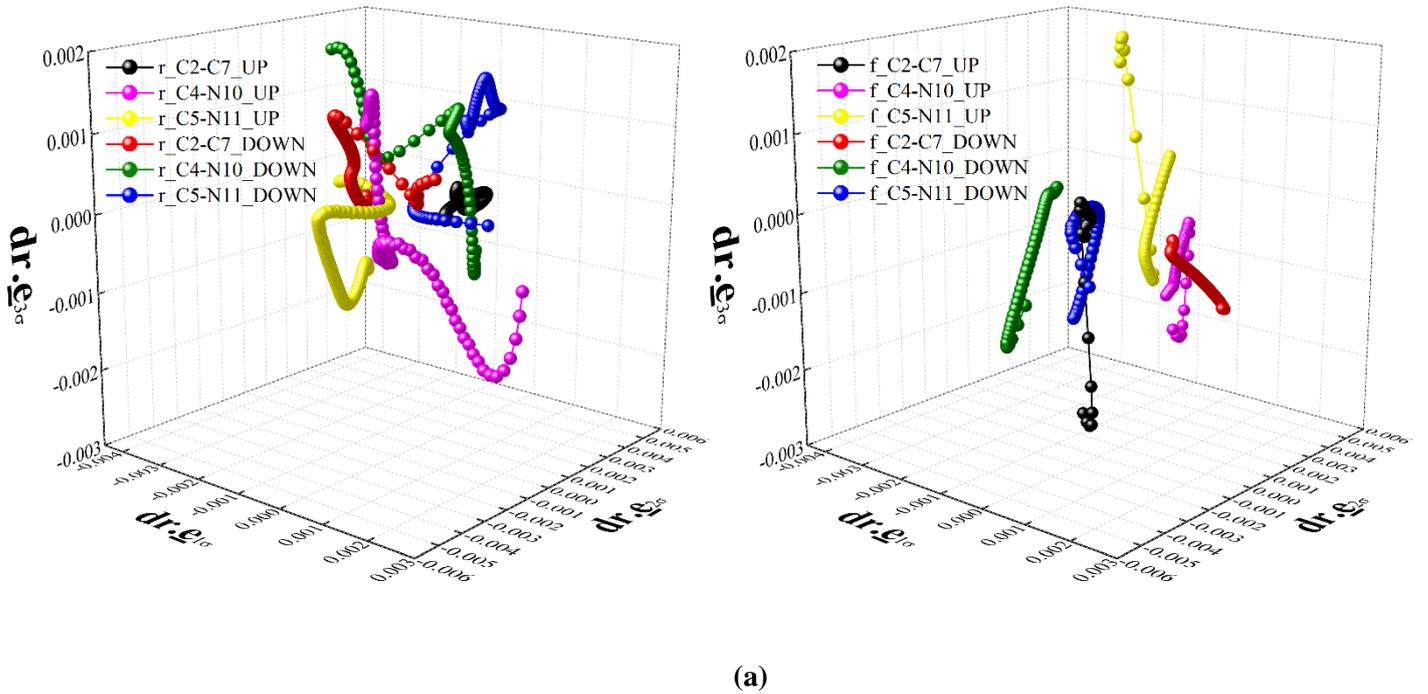

**(a)**

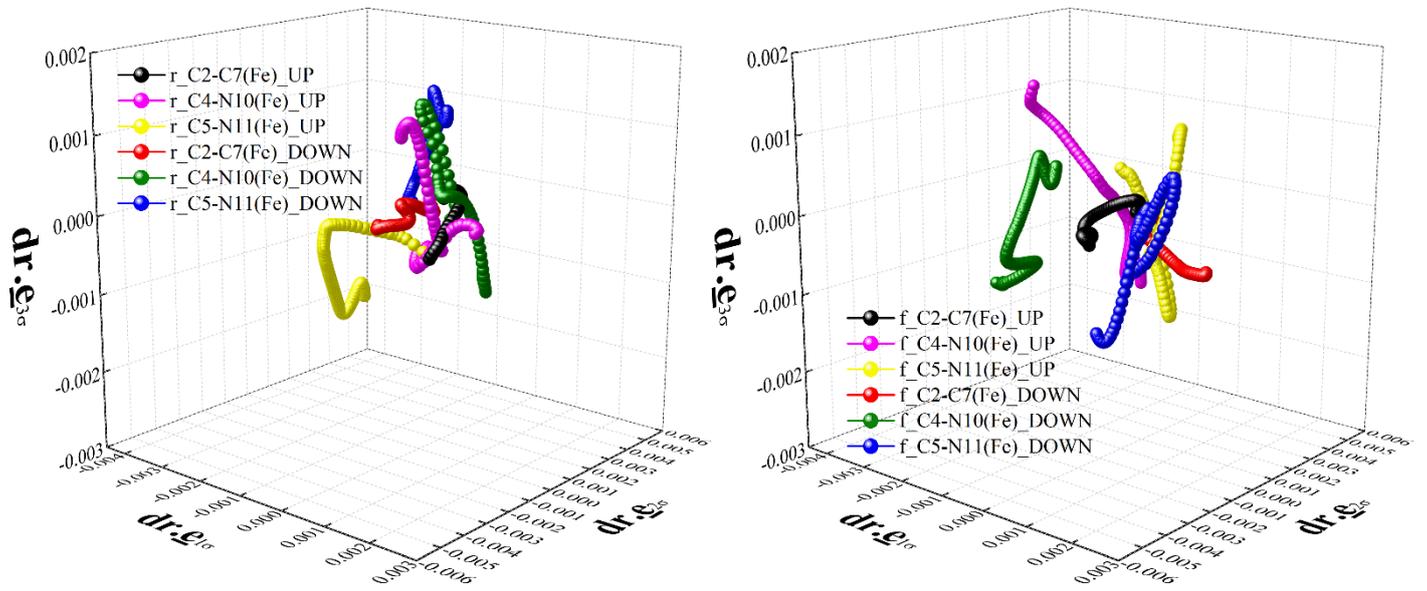

**(b)**

**Figure 6**. The trajectories $\mathbb{T}_\sigma(s)$ in the eigenvector projection space $\mathbb{U}_\sigma(s)$ for C2-C7 *BCP*, C4-N10 *BCP* and C5-N11 *BCP* of the F decorated switch without and with Fe are shown in sub-figures **(a)** and **(b),** respectively.

It can be seen that for both the undoped and Fe-doped quinone switch, the UP and DOWN trajectories $\mathbb{T}_\sigma(s)$ for the N11-C5 ('entry' pathway) and C2-C7 *BCP* ('exit' pathway) trace very different paths through $\mathbb{U}_\sigma$ space and are therefore distinct and unique, see **Figure 6(a-b)** respectively. This is in contrast to the scalar results for the ellipticity ε, total local energy density $H(\mathbf{r_b})$ and the stress tensor eigenvalue $\lambda_{3\sigma}$, see **Figure 1(a-b)**, **Figure 3(a-b)**, **Figure 4(a-b)** and **Figure 5(a-b)** respectively.

Differences between the undoped and Fe-doped quinone switch are apparent from the lack of symmetry, for the undoped switch between the reverse (*r*) 'ON' position and the forward (*f*) 'OFF' position, see **Figure 6(a).** In particular, it is obvious that the trajectories $\mathbb{T}_\sigma(s)$ are considerably more linear in character towards the forward (*f*) 'OFF' position than the reverse (*r*) 'ON' position. Conversely, for the Fe-doped switch there is a preservation of symmetry between the reverse (*r*) 'ON' position and the forward (*f*) 'OFF' position in keeping with Hammond's postulate, see **Figure 6(b)**[30].

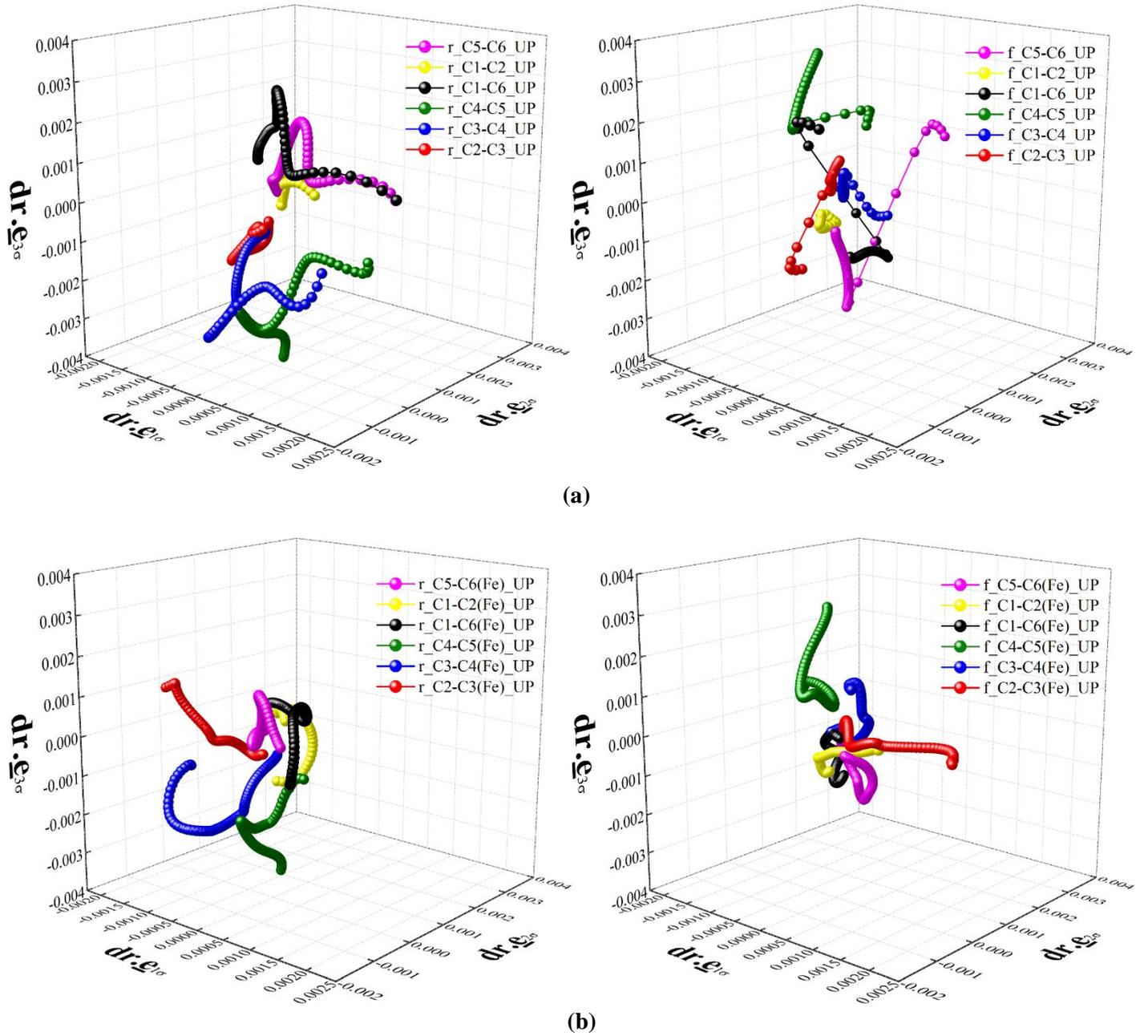

**Figure 7**. The trajectories $\mathbb{T}_\sigma(s)$ in the eigenvector projection space $\mathbb{U}_\sigma(s)$ for the C-C *BCP*s with no Fe and Fe are shown in sub-figures **(a)** and **(b)** respectively.

The effect of the addition of an iron atom to the quinone switch is seen in the mixing of three C-C trajectories $\mathbb{T}_\sigma(s)$, corresponding to the C4-C5 *BCP*, C3-C4 *BCP* and the C2-C3 *BCP*s comprising the *North* pathway (C5-C4-C3-C2) and C5-C6 *BCP*, C1-C2 *BCP* and the C1-C6 *BCP*s comprising the *South* pathway (C5-C6-C1-C2); compare the undoped switch C-C trajectories $\mathbb{T}_\sigma(s)$ in **Figure 7(a)** with the Fe-doped C-C trajectories $\mathbb{T}_\sigma(s)$ in **Figure 7(b)**. The *North* pathway (C5-C4-C3-C2) and *South* pathway (C5-C6-C1-C2) for the C-C trajectories $\mathbb{T}_\sigma(s)$ of the undoped switch are clearly separated, unlike those for the Fe-doped switch.

This reflects the results for the ellipticity ε, total local energy density $H(r_b)$ and the stress tensor eigenvalue $\lambda_{3\sigma}$, see **Figure 3(c-d)**, **Figure 4(c-d)** and **Figure 5(c-d)** respectively.

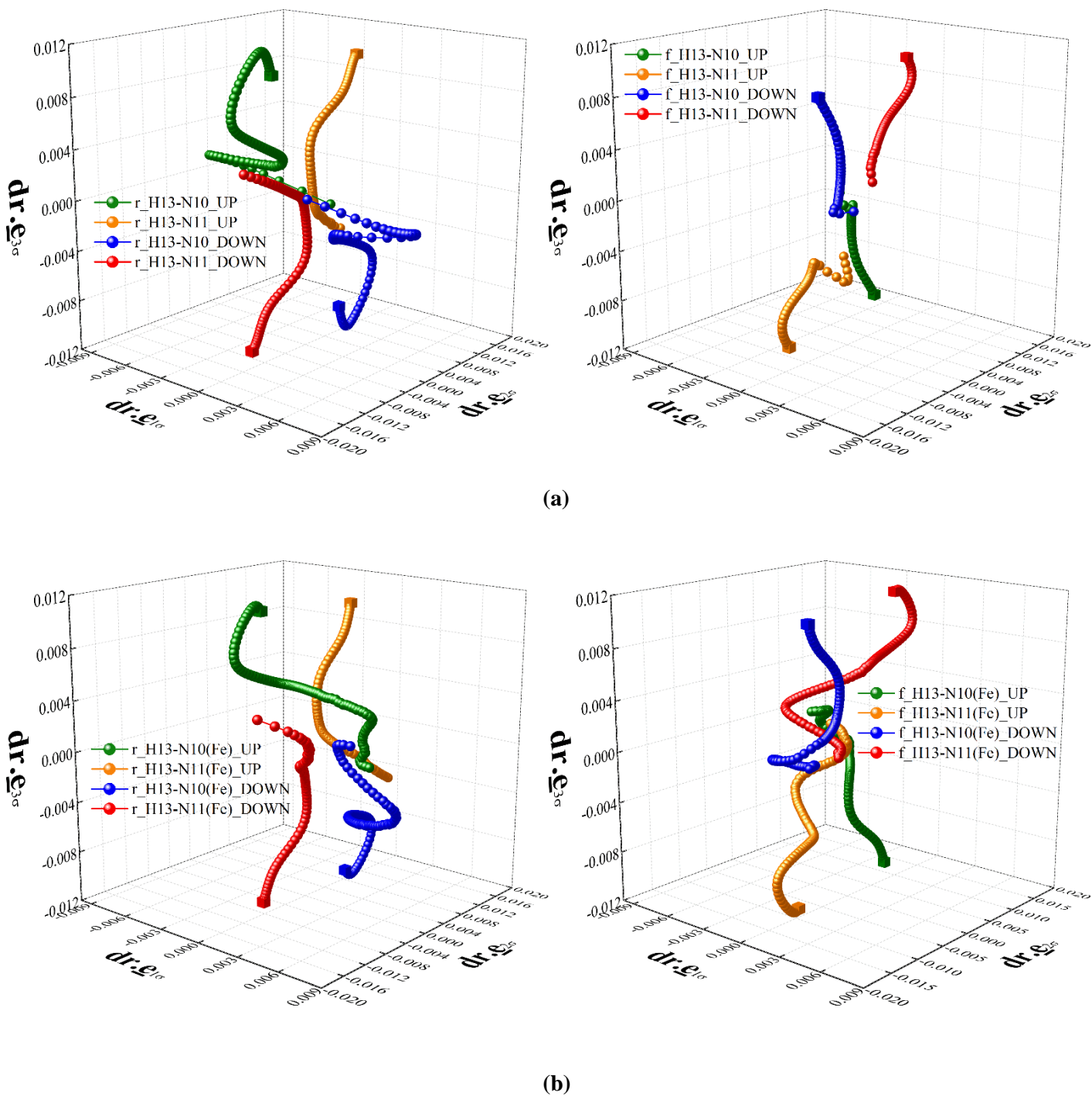

**Figure 8**. The trajectories $\mathbb{T}_\sigma(s)$ in the eigenvector projection space $\mathbb{U}_\sigma(s)$ for the N-H *BCP*s of F with no Fe and Fe are shown in sub-figures **(a)** and **(b)** respectively.

The *BCP* associated with the hydrogen *NCP* motion between the N10 and N11 *NCP*s moves considerably further in both real space and $\mathbb{U}_\sigma$ space than all other *BCP*s, see **Figure 8(a-b)**. This is determined by the lengths of the trajectories $\mathbb{T}_\sigma(s)$ specified in real space by *l* and in $\mathbb{U}_\sigma$ space by $\mathbb{L}_\sigma$, see **Table 1** and **Table 2**. There is a strong contrast in the form of the trajectories $\mathbb{T}_\sigma(s)$ of the H13--N10/N10-H13 *BCP* and

H13--N11/N11-H13 *BCP* for the forward 'OFF' switch position between the undoped and Fe-doped quinone switches, see **Figure 8(a)** and **Figure 8(b)** respectively. The linear trajectories $\mathbb{T}_\sigma(s)$ of the forward 'OFF' switching in the undoped switch resembles the form of the trajectories $\mathbb{T}_\sigma(s)$ for fatigue reaction recently examined[21], see **Figure 8(a)** whereas the validity of Hammond's postulate is again apparent for the Fe-doped switch, see **Figure 8(b)**.

The greater symmetry of the Fe-doped switch between the reverse (*r*) 'ON' and forward (*f*) 'OFF' trajectories $\mathbb{T}_\sigma(s)$ for the N10-H13 *BCP* compared with the undoped switch is apparent in the lengths (*l* in real space and $\mathbb{L}_\sigma$ in $\mathbb{U}_\sigma$ space), compare **Table 2** with **Table 1.** There is also a more equal extent along both the most preferred $\underline{e}_{1\sigma}$ and the least preferred direction $\underline{e}_{2\sigma}$ for the Fe-doped switch, see **Figure 8(b)** and also compare the entries for reverse (*r*) 'ON' and forward (*f*) 'OFF' of the N10-H13 *BCP* in **Table 3** and **Table 4**. The undoped switch corresponding to the $\mathbb{T}_\sigma(s)$ of the H13--N10/N10-H13 *BCP* does not attain the 'OFF' position in contrast to the corresponding $\mathbb{T}_\sigma(s)$ of the Fe-doped switch. This observation is explained by the existence of the significantly shorter $\mathbb{T}_\sigma(s)$ in the most preferred direction $\underline{e}_{1\sigma}$ of the undoped switch in the 'OFF' position compared with the Fe-doped switch; compare **Table 3** with **Table 4.**

**Table 1**. The stress tensor projection space $\mathbb{U}_\sigma$ trajectory lengths $\mathbb{L}_\sigma$ in atomic units (a.u.) for the reverse (*r*) and forward (*f*) directions of the IRC of the F *NCP* substituent without the Fe *NCP* dopant, C3-F8 (UP position) and C6-F8 (DOWN position) refer to the positions of the F substituent indicated in **Figure 1(a)** and **Figure 1(b)** respectively.

|  | $\mathbb{L}_\sigma$ | | *l* | |
|---|---|---|---|---|
|  | Up | Down | Up | Down |
| BCP | (*r*,*f*) | (*r*,*f*) | (*r*,*f*) | (*r*,*f*) |
| C1-C2 | (0.0039, 0.0012) | (0.0023, 0.0016) | (0.1182, 0.0204) | (0.0703, 0.0368) |
| C1-C6 | (0.0060, 0.0066) | (0.0037, 0.0018) | (0.2364, 0.0844) | (0.1351, 0.0686) |
| C2-C3 | (0.0029, 0.0038) | (0.0057, 0.0029) | (0.1015, 0.0442) | (0.1508, 0.0847) |
| C2-C7 | (0.0054, 0.0052) | (0.0070, 0.0033) | (0.2802, 0.0912) | (0.2637, 0.1437) |
| C3-C4 | (0.0063, 0.0031) | (0.0071, 0.0032) | (0.2241, 0.0305) | (0.2267, 0.0697) |
| C4-C5 | (0.0065, 0.0047) | (0.0067, 0.0029) | (0.3142, 0.1275) | (0.3126, 0.1349) |
| C5-C6 | (0.0052, 0.0074) | (0.0039, 0.0033) | (0.1399, 0.0878) | (0.0674, 0.0907) |
| C1-N14 | (0.0061, 0.0070) | (0.0052, 0.0027) | (0.2982, 0.1059) | (0.1919, 0.1049) |
| C4-N10 | (0.0084, 0.0035) | (0.0078, 0.0036) | (0.2406, 0.1455) | (0.2374, 0.1731) |
| C5-N11 | (0.0076, 0.0059) | (0.0076, 0.0038) | (0.3622, 0.0770) | (0.3091, 0.0794) |
| H13-N10 | (0.0421, 0.0128) | (0.0397, 0.0147) | (1.0105, 0.1630) | (1.0618, 0.1720) |
| H13-N11 | (0.0283, 0.0190) | (0.0282, 0.0143) | (0.5249, 0.4782) | (0.5068, 0.4619) |
| H17-N10 | (0.0419, 0.0192) | (0.0363, 0.0079) | (0.7118, 0.4388) | (0.7323, 0.4850) |
| H12-N11 | (0.0247, 0.0150) | (0.0250, 0.0121) | (0.9841, 0.1526) | (0.8469, 0.1523) |
| C3-F8 | (0.0064, 0.0057) | ( ---, --- ) | (0.2813, 0.0770) | ( ---, --- ) |
| C6-F9 | ( ---, --- ) | (0.0056, 0.0030) | ( ---, --- ) | (0.1769, 0.1271) |

**Table 2**. The stress tensor projection space $\mathbb{U}_\sigma$ trajectory lengths $\mathbb{L}_\sigma$ in atomic units (a.u.) of the F *NCP* substituent for the Fe *NCP* dopant, see the caption of **Table 1** for further details.

|  | $\mathbb{L}_\sigma$ | | *l* | |
|---|---|---|---|---|
|  | *Up* | *Down* | *Up* | *Down* |
| *BCP* | (*r,f*) | (*r,f*) | (*r,f*) | (*r,f*) |
| C1-C2 | (0.0031, 0.0018) | (0.0023, 0.0031) | (0.0852, 0.0694) | (0.0589, 0.0991) |
| C1-C6 | (0.0038, 0.0026) | (0.0025, 0.0026) | (0.1172, 0.0945) | (0.0671, 0.0723) |
| C2-C3 | (0.0028, 0.0029) | (0.0036, 0.0045) | (0.1051, 0.1360) | (0.1234, 0.1806) |
| C2-C7 | (0.0045, 0.0032) | (0.0036, 0.0058) | (0.1498, 0.1291) | (0.1230, 0.1814) |
| C3-C4 | (0.0046, 0.0028) | (0.0045, 0.0035) | (0.1818, 0.1067) | (0.1807, 0.1237) |
| C4-C5 | (0.0048, 0.0051) | (0.0043, 0.0052) | (0.2340, 0.2322) | (0.2325, 0.2188) |
| C5-C6 | (0.0034, 0.0035) | (0.0027, 0.0041) | (0.0735, 0.1482) | (0.0723, 0.1417) |
| C1-N14 | (0.0047, 0.0027) | (0.0032, 0.0043) | (0.1644, 0.1327) | (0.1128, 0.1240) |
| C4-N10 | (0.0053, 0.0051) | (0.0063, 0.0054) | (0.1366, 0.3553) | (0.1322, 0.3218) |
| C5-N11 | (0.0050, 0.0063) | (0.0036, 0.0065) | (0.2895, 0.1566) | (0.2736, 0.1378) |
| H13-N10 | (0.0274, 0.0249) | (0.0310, 0.0259) | (1.0099, 0.6132) | (1.0161, 0.5948) |
| H13-N11 | (0.0244, 0.0285) | (0.0281, 0.0297) | (0.5262, 0.9107) | (0.5650, 0.9108) |
| H17-N10 | (0.0262, 0.0186) | (0.0336, 0.0180) | (0.6609, 0.9327) | (0.7490, 0.8776) |
| H12-N11 | (0.0192, 0.0378) | (0.0153, 0.0383) | (0.8044, 0.7452) | (0.7496, 0.7074) |
| C3-Fe18 | (0.0035, 0.0029) | (0.0042, 0.0043) | (0.1282, 0.1338) | (0.1426, 0.1587) |
| C6-Fe18 | (0.0029, 0.0042) | (0.0028, 0.0039) | (0.0628, 0.1767) | (0.0447, 0.1234) |
| C3-F8 | (0.0062, 0.0045) | ( --- , --- ) | (0.2456, 0.2448) | ( --- , --- ) |
| C6-F9 | ( --- , --- ) | (0.0039, 0.0054) | ( --- , --- ) | (0.1201, 0.1855) |

**Table 3.** The maximum projections $\{(\mathbf{e}_{1\sigma}\cdot d\mathbf{r})_{max}, (\mathbf{e}_{2\sigma}\cdot d\mathbf{r})_{max}, (\mathbf{e}_{3\sigma}\cdot d\mathbf{r})_{max}\}$ in the stress tensor eigenvector projection space $\mathbb{U}_\sigma$ of the F *NCP* substituent without the Fe *NCP* dopant are calculated using the eigenvectors, see the main text for further details.

| BCP | UP reverse $\{(\mathbf{e}_{1\sigma}\cdot d\mathbf{r})_{max}, (\mathbf{e}_{2\sigma}\cdot d\mathbf{r})_{max}, (\mathbf{e}_{3\sigma}\cdot d\mathbf{r})_{max}\}$ | UP forward $\{(\mathbf{e}_{1\sigma}\cdot d\mathbf{r})_{max}, (\mathbf{e}_{2\sigma}\cdot d\mathbf{r})_{max}, (\mathbf{e}_{3\sigma}\cdot d\mathbf{r})_{max}\}$ | DOWN reverse $\{(\mathbf{e}_{1\sigma}\cdot d\mathbf{r})_{max}, (\mathbf{e}_{2\sigma}\cdot d\mathbf{r})_{max}, (\mathbf{e}_{3\sigma}\cdot d\mathbf{r})_{max}\}$ | DOWN forward $\{(\mathbf{e}_{1\sigma}\cdot d\mathbf{r})_{max}, (\mathbf{e}_{2\sigma}\cdot d\mathbf{r})_{max}, (\mathbf{e}_{3\sigma}\cdot d\mathbf{r})_{max}\}$ |
|---|---|---|---|---|
| N10-H13 | {8.1501E-03, 1.3946E-02, 1.1974E-02} | {4.4898E-04, 5.6062E-03, 8.7262E-03} | {8.2269E-03, 1.4298E-02, 1.1961E-02} | {3.0265E-04, 5.6332E-03, 8.3910E-03} |
| N11-H13 | {1.8487E-03, 1.2140E-02, 1.0356E-02} | {5.6866E-04, 1.2803E-02, 1.0044E-02} | {1.2838E-04, 1.1839E-02, 1.0645E-02} | {4.5642E-04, 1.2707E-02, 9.9879E-03} |
| C3--F8 | {7.4705E-04, 3.3810E-03, 1.8788E-03} | {3.2599E-04, 2.1697E-03, 3.6119E-04} | { -  ,  -  ,  - } | { - ,  - ,  - } |
| C6--F9 | { - ,  - ,  - } | { - ,  - ,  - } | {9.0101E-04, 2.0959E-03, 4.7423E-04} | {5.8482E-05, 3.7263E-03, 1.7540E-03} |
| C4-N10 | {2.6622E-03, 2.4184E-03, 1.6407E-03} | {1.2894E-04, 3.5603E-03, 1.8932E-03} | {2.4812E-03, 2.5866E-03, 1.8729E-03} | {2.4689E-04, 4.5647E-03, 1.2133E-03} |
| C5-N11 | {1.2202E-03, 4.5080E-03, 7.7812E-04} | {2.3003E-04, 2.2722E-03, 2.1593E-03} | {1.0667E-03, 3.8115E-03, 1.5432E-03} | {2.4447E-04, 2.3160E-03, 1.0991E-03} |
| C4-C5 | {1.7123E-03, 8.8008E-04, 3.9426E-03} | {2.2434E-04, 1.3265E-03, 3.8355E-03} | {1.0764E-03, 1.6392E-03, 3.9169E-03} | {9.3235E-05, 4.7950E-04, 3.8094E-03} |
| C4-C3 | {1.2551E-03, 1.5706E-03, 2.7655E-03} | {9.2429E-05, 1.3819E-03, 8.7114E-04} | {3.0095E-04, 1.1093E-03, 3.3405E-03} | {8.8565E-04, 3.1636E-04, 2.8321E-03} |
| C3-C2 | {2.8275E-04, 6.6110E-04, 1.1923E-03} | {1.4000E-04, 9.8032E-04, 1.3111E-03} | {1.0231E-03, 1.7611E-03, 1.4403E-03} | {7.2019E-04, 9.1056E-04, 2.9061E-03} |
| C2-C7 | {2.7953E-04, 3.6747E-03, 2.1789E-04} | {7.8771E-04, 2.1954E-03, 2.7662E-03} | {1.0363E-03, 3.3882E-03, 1.4468E-03} | {5.4125E-04, 4.3601E-03, 1.5568E-03} |
| C2-C1 | {6.7637E-04, 5.8036E-04, 1.7712E-03} | {2.1995E-04, 3.5204E-04, 4.3596E-04} | {1.8486E-04, 4.3100E-04, 8.3285E-04} | {8.4771E-05, 9.5757E-04, 1.2141E-03} |
| C1-C6 | {2.3712E-03, 3.4663E-04, 3.0118E-03} | {4.5831E-04, 1.4885E-03, 2.4354E-03} | {7.5800E-04, 2.8614E-04, 1.6002E-03} | {3.4673E-04, 9.7967E-04, 1.6496E-03} |
| C6-C5 | {2.1503E-03, 2.5534E-04, 2.2235E-03} | {5.7470E-04, 2.5404E-03, 2.6336E-03} | {5.8309E-04, 8.2704E-04, 1.1267E-03} | {2.3876E-04, 1.4890E-03, 2.8112E-03} |

**Table 4.** The maximum projections {$(\underline{e}_{1\sigma}\cdot\mathbf{dr})_{max}$, $(\underline{e}_{2\sigma}\cdot\mathbf{dr})_{max}$, $(\underline{e}_{3\sigma}\cdot\mathbf{dr})_{max}$} in the stress tensor eigenvector projection space $\mathbb{U}_\sigma$ of the F *NCP* substituent for the Fe *NCP* dopant are calculated using the eigenvectors, see the main text for further details.

| | UP | | DOWN | |
|---|---|---|---|---|
| | *reverse* | *forward* | *reverse* | *forward* |
| BCP | {$(e_{1\sigma}\cdot dr)_{max}$, $(e_{2\sigma}\cdot dr)_{max}$, $(e_{3\sigma}\cdot dr)_{max}$} | {$(e_{1\sigma}\cdot dr)_{max}$, $(e_{2\sigma}\cdot dr)_{max}$, $(e_{3\sigma}\cdot dr)_{max}$} | {$(e_{1\sigma}\cdot dr)_{max}$, $(e_{2\sigma}\cdot dr)_{max}$, $(e_{3\sigma}\cdot dr)_{max}$} | {$(e_{1\sigma}\cdot dr)_{max}$, $(e_{2\sigma}\cdot dr)_{max}$, $(e_{3\sigma}\cdot dr)_{max}$} |
| N10-H13 | {5.4545E-03,1.3646E-02,1.1819E-02} | {3.6783E-03,7.8036E-03,1.0487E-02} | {6.1868E-03,1.3772E-02,1.1886E-02} | {4.4265E-03,7.7174E-03,1.0377E-02} |
| N11-H13 | {2.5403E-03,1.0907E-02,1.0689E-02} | {3.5148E-03,1.3506E-02,1.1525E-02} | {3.1837E-03,1.3164E-02,1.0877E-02} | {4.9805E-03,1.3446E-02,1.1556E-02} |
| C3--F8 | {3.8539E-03,1.6718E-03,1.4150E-03} | {3.1802E-03,1.5968E-03,3.4224E-04} | { - , - , - } | { - , - , - } |
| C6--F9 | { - , - , - } | { - , - , - } | {1.8454E-03,1.4846E-03,6.9366E-04} | {1.6778E-03,1.8422E-03,1.0911E-03} |
| | | | | |
| C4-N10 | {1.5076E-03,1.9812E-03,1.3471E-03} | {3.1080E-03,2.4803E-03,1.3005E-03} | {7.4327E-04,2.1341E-03,1.3216E-03} | {2.3678E-03,3.1186E-03,1.0377E-03} |
| C5-N11 | {5.7336E-04,3.2418E-03,1.0427E-03} | {9.2856E-04,2.0488E-03,1.1615E-03} | {1.4189E-03,2.6177E-03,1.3695E-03} | {1.3241E-03,1.8326E-03,1.3474E-03} |
| C4-C5 | {6.0775E-04,9.6975E-04,3.3424E-03} | {1.1353E-03,1.0821E-03,3.3401E-03} | {9.6410E-04,1.2212E-03,3.3243E-03} | {3.1371E-04,8.8207E-04,3.3210E-03} |
| C4-C3 | {1.1088E-03,1.5914E-03,1.8723E-03} | {5.0500E-04,4.1581E-04,1.5420E-03} | {2.0661E-03,1.2677E-03,1.8049E-03} | {1.1381E-03,2.8891E-04,1.4671E-03} |
| C3-C2 | {1.6781E-03,5.8772E-04,1.2112E-03} | {1.9699E-03,3.5055E-04,6.4300E-04} | {2.3350E-03,8.0272E-04,1.0719E-03} | {2.8000E-03,6.0125E-04,1.0938E-03} |
| C2-C7 | {1.0291E-03,1.7573E-03,2.9220E-04} | {4.3703E-04,1.8394E-03,2.9242E-04} | {3.2753E-04,1.9343E-03,2.3793E-04} | {2.1565E-03,1.9150E-03,4.3399E-04} |
| C2-C1 | {3.5227E-04,5.9869E-04,1.0428E-03} | {7.0042E-04,2.4601E-04,7.7386E-04} | {3.9578E-04,5.3324E-04,8.4333E-04} | {1.3915E-03,3.4412E-04,7.6561E-04} |
| C1-C6 | {9.6714E-04,8.0913E-04,1.1037E-03} | {4.9324E-04,7.1875E-04,1.0513E-03} | {7.7863E-04,8.9428E-04,7.7369E-04} | {4.1381E-04,3.9723E-04,7.0172E-04} |
| C6-C5 | {6.2487E-04,6.7077E-04,1.3444E-03} | {4.0670E-04,7.4140E-04,1.6384E-03} | {4.4546E-04,7.3303E-04,1.1025E-03} | {3.1377E-04,1.0947E-03,1.5370E-03} |
| C3-Fe18 | {1.0419E-03,9.0710E-04,2.2176E-03} | {7.1612E-04,7.6344E-04,1.7666E-03} | {1.3334E-03,1.2788E-03,2.9402E-03} | {8.2297E-04,9.1599E-04,2.3458E-03} |
| C6-Fe18 | {9.9905E-04,5.9118E-04,9.2609E-04} | {7.1787E-04,2.5634E-03,4.4577E-04} | {1.2893E-03,7.0223E-04,5.4018E-04} | {7.9916E-04,1.8633E-03,3.7942E-04} |

## Conclusions

Using both a QTAIM and stress tensor analysis we have explained how the proton transfer mechanism responsible for the functioning of quinone switches is affected by the absence or presence of a dopant Fe *NCP*. The findings from the QTAIM and stress tensor scalar measures included observation of increasing ellipticity ε values and a 'spreading out' and mixing of differences in the ellipticity ε that is associated with increased ease of conductivity when the Fe *NCP* is present. The Fe *NCP* increased the ellipticity ε of the N11-C5 *BCP* ('entry' pathway), reduced the *BCP* strength on the basis of less negative values of the total local energy density $H(\mathbf{r_b})$ but increased *BCP* stability based on less negative values of the stress tensor eigenvalue $\lambda_{3\sigma}$. In doing so we also demonstrated that the desired switch properties were more pronounced when an F *NCP* was placed in the UP position compared with the DOWN position, although this effect was secondary to the presence of the Fe *NCP*. The absence of the Fe-dopant *NCP* coincided with the formation of closed-shell H---H *BCP*s that were not present in the 'OFF' position for the Fe-doped quinone switch. The closed shell H---H *BCP*s indicate a strained or electron deficient environment that also occurs in other systems such as biphenyl[5].

The main finding from the stress tensor analysis was that the introduction of the Fe *NCP* into the switch increased the symmetry and hence the desired functioning of the switch as determined by the trajectories $\mathbb{T}_\sigma(s)$ in the $\mathbb{U}_\sigma$ space and in doing so we rediscovered Hammond's postulate. The addition of the Fe *NCP* caused the removal of the separation of the trajectories $\mathbb{T}_\sigma(s)$ of the *North* pathway or *South* pathways in a manner indicating the conditions for increased conductivity, consistent with the results from the ellipticity ε. We found the trajectories $\mathbb{T}_\sigma(s)$ to be particularly suited to following the proton transfer reaction due to the large distances moved by the H13--N11/N11-H13 *BCP* and the H13--N10/N10-H13 *BCP*. The trajectories $\mathbb{T}_\sigma(s)$ of the H13--N11/N11-H13 *BCP* and the H13--N10/N10-H13 *BCP*s provide visualizations in $\mathbb{U}_\sigma$ space of the 'ON' and "OFF" switch mechanisms respectively. The trajectories $\mathbb{T}_\sigma(s)$ can follow the process of the switch in the 'ON' and 'OFF' position because the trajectories $\mathbb{T}_\sigma(s)$ are defined to start at the transition state and end at the reverse (*r*) and forward (*f*) minimum, respectively, rather than starting at the minima and ending at the ambiguous transition state.

We find that the undoped quinone switch does not fully attain the 'OFF' position due to the lack of extent of the $\mathbb{T}_\sigma(s)$ of the H13--N10/N10-H13 *BCP* in the $\underline{\mathbf{e}}_{1\sigma}$ direction, i.e. the easiest direction of electronic motion. The consequence of this will be a leaky switch which makes the molecule less suitable to being used in electronic circuits. This problem is corrected by the addition of the Fe-dopant atom explained by significant motion of the H13--N10/N10-H13 *BCP* in the preferred $\underline{\mathbf{e}}_{1\sigma}$ direction and therefore the attainment of the 'OFF' position in the $\mathbb{U}_\sigma$ space representation.

The insights gained from this stress tensor analysis have been demonstrated to be more useful than a conventional Cartesian-based analysis for exploration of the properties of molecular electronic devices. Future investigations could consider additional dopants or the application of electric fields that enable the switch to function optimally by readily moving between the 'ON' and 'OFF' positions.

**Acknowledgements**

The National Natural Science Foundation of China is acknowledged, project approval number: 21673071. The One Hundred Talents Foundation of Hunan Province and the aid program for the Science and Technology Innovative Research Team in Higher Educational Institutions of Hunan Province are gratefully acknowledged for the support of S.J. and S.R.K. The Royal Society is thanked by S.J., S.R.K, T.X, T.v.M and H.F. for support through an International Exchanges grant. We thank EaStCHEM for computational support via the EaStCHEM Research Computing Facility.